\pdfoutput=1
\documentclass[sigconf]{acmart}

\newif\ifanonymous
\anonymousfalse

\AtBeginDocument{%
  }

\copyrightyear{2026}
\acmYear{2026}
\setcopyright{cc}
\setcctype{by}
\acmConference[VRST '26]{32nd ACM Symposium on Virtual Reality Software and Technology}{November 16--18, 2026}{Sendai, Japan}
\acmBooktitle{32nd ACM Symposium on Virtual Reality Software and Technology (VRST '26), November 16--18, 2026, Sendai, Japan}
\acmDOI{10.1145/3822517.3848668}
\acmISBN{979-8-4007-2811-2/2026/11}

\newcommand{\thesystem}{\textsc{Vibe Coding XR}\xspace}
\newcommand{\xrblocks}{\textsc{XR Blocks}\xspace}
\newcommand{\vcxr}{\textsc{vcxr60}\xspace}

\ifanonymous
    \newcommand{\citexrblocks}{}
    \newcommand{\xburl}{\textit{[URL omitted]\xspace}}
    \newcommand{\gemurl}{\textit{[URL omitted]\xspace}}
    \newcommand{\prompturl}{\textit{[URL omitted]\xspace}}
\else
    \newcommand{\citexrblocks}{\cite{Li2025XR}}
    \newcommand{\xburl}{\url{https://github.com/google/xrblocks}}
    \newcommand{\gemurl}{\url{http://xrblocks.github.io/gem}}
    \newcommand{\prompturl}{\url{http://xrblocks.github.io/prompts}}
\fi

\usepackage[utf8]{inputenc}
\usepackage{amsmath}
\usepackage{enumitem}
\usepackage{subfigure}
\usepackage{fontawesome5}
\usepackage{xspace}
\usepackage{xcolor}
\usepackage{listings}
\usepackage{makecell}
\usepackage{bbding}

\lstdefinelanguage{Markdown}{
    morekeywords={\#}, 
    sensitive=false,
    morecomment=[l]{\#},
    morestring=[b]",
    morestring=[b]',
    moredelim=[is][\color{blue}\bfseries]{**}{**},
    moredelim=[is][\color{teal}\ttfamily]{`}{`},
    alsoletter={\#}
}

\lstdefinelanguage{JavaScriptXR}{
    morekeywords={async,await,break,case,catch,class,const,constructor,continue,
        default,delete,do,else,export,extends,finally,for,from,function,if,
        import,in,instanceof,let,new,return,super,switch,this,throw,try,
        typeof,var,void,while,yield,true,false,null,undefined},
    sensitive=true,
    morecomment=[l]{//},
    morecomment=[s]{/*}{*/},
    morestring=[b]',
    morestring=[b]",
    morestring=[b]`,
}
\lstdefinestyle{jsstyle}{
    language=JavaScriptXR,
    basicstyle=\ttfamily\footnotesize,
    keywordstyle=\color{teal!70!black}\bfseries,
    commentstyle=\color{gray},
    stringstyle=\color{red!60!black},
    backgroundcolor=\color{lightgray!10},
    breaklines=true,
    breakatwhitespace=false,
    columns=fullflexible,
    keepspaces=true,
    showstringspaces=false,
    upquote=true,
    frame=none,
    numbers=none,
    xleftmargin=0pt,
    aboveskip=6pt,
    belowskip=6pt,
}
\lstnewenvironment{jscode}{\lstset{style=jsstyle}}{}

\usepackage{tcolorbox}

\usepackage{multicol}
\usepackage{comment}
\usepackage{booktabs,dcolumn}
\usepackage{multirow}
\usepackage{threeparttable}
\usepackage{siunitx} 
\usepackage{fancyhdr}	

\newcolumntype{L}[1]{>{\raggedright\let\newline\\\arraybackslash\hspace{0pt}}m{#1}}
\newcolumntype{C}[1]{>{\centering\let\newline\\\arraybackslash\hspace{0pt}}m{#1}}
\newcolumntype{R}[1]{>{\raggedleft\let\newline\\\arraybackslash\hspace{0pt}}m{#1}}

\renewenvironment{quote}[1][0.04\linewidth]
  {\list{}{\leftmargin=#1\rightmargin=#1}\item\relax}{\endlist}

\usepackage[normalem]{ulem}

\input{_author_hack}

\begin{document}
\title[\thesystem]{Vibe Coding XR: Accelerating AI + XR Prototyping \protect\\with XR Blocks and Gemini}

\settopmatter{printacmref=false}

\author{Ruofei Du}
\authornote{Both authors contributed equally to XR Blocks v0.1 -- v0.11. We thank latest XR Blocks contributors on GitHub, especially v0.20+ major improvements by William Huang.}
\authornote{Corresponding author: \href{mailto:me@duruofei.com}{me@duruofei.com}.}

\author{Benjamin Hersh}

\author{David Li}
\authornotemark[1]

\author{Nels Numan}
\authornote{Equal contribution to XR Blocks, sorted alphabetically.}

\author{Xun Qian}
\authornotemark[3]

\author{Yanhe Chen}
\authornotemark[3]

\author{Zhongyi Zhou}
\authornotemark[3]

\author{Jiahao Ren}

\author{Xingyue Chen}

\author{Robert Timothy Bettridge}

\author{Faraz Faruqi}

\author{Xiang `Anthony' Chen}

\author{Steve Toh}

\author{David Kim}
\affiliation{
    \country{ Try at Google AI Studio: \url{https://goo.gle/BuildXR} or XR Blocks Gem: \url{https://xrblocks.github.io/gem} \\
    \href{https://github.com/google/xrblocks}{\faGitSquare}  \url{https://github.com/google/xrblocks} \\
    }
    \institution{\textbf{Google XR Labs}}
}

\renewcommand{\shortauthors}{Du et al.}

\begin{abstract}
While large language models (LLMs) have accelerated traditional software development through intent-driven ``vibe coding'', prototyping intelligent Extended Reality (XR) experiences remains a major challenge. The fundamental barrier is not just the steep learning curve for human creators, but that low-level sensor APIs and complex game engine hierarchies are ill-suited for LLM reasoning, routinely exceeding context windows and inducing syntax hallucinations. To bridge this gap, we contribute \xrblocks, an open-source, \textit{LLM-native} WebXR framework. Unlike traditional engines, \xrblocks introduces a semantic ``Reality Model'' that aligns spatial computing primitives (users, physical environments, and agents) with natural language, providing a robust, concise vocabulary optimized for generative AI. Building upon this foundation, we present \thesystem, an end-to-end prototyping workflow that leverages LLMs to translate high-level prompts (\textit{e.g.}, \textit{``create a dandelion that reacts to my hand''}) directly into functional, physics-aware mixed-reality applications. To minimize the friction of on-device testing, the workflow introduces a seamless desktop ``simulated reality'' to headset deployment loop. Finally, we introduce \vcxr, a pilot dataset of 60 XR prompts paired with an automated evaluation pipeline. Our technical evaluation demonstrates high one-shot execution success, enabling practitioners to bypass low-level hurdles and rapidly move from ``idea to reality.'' Code and demo are available at \xburl\xspace and \url{https://goo.gle/BuildXR}.
\end{abstract}



\begin{teaserfigure}
  \includegraphics[width=\textwidth]{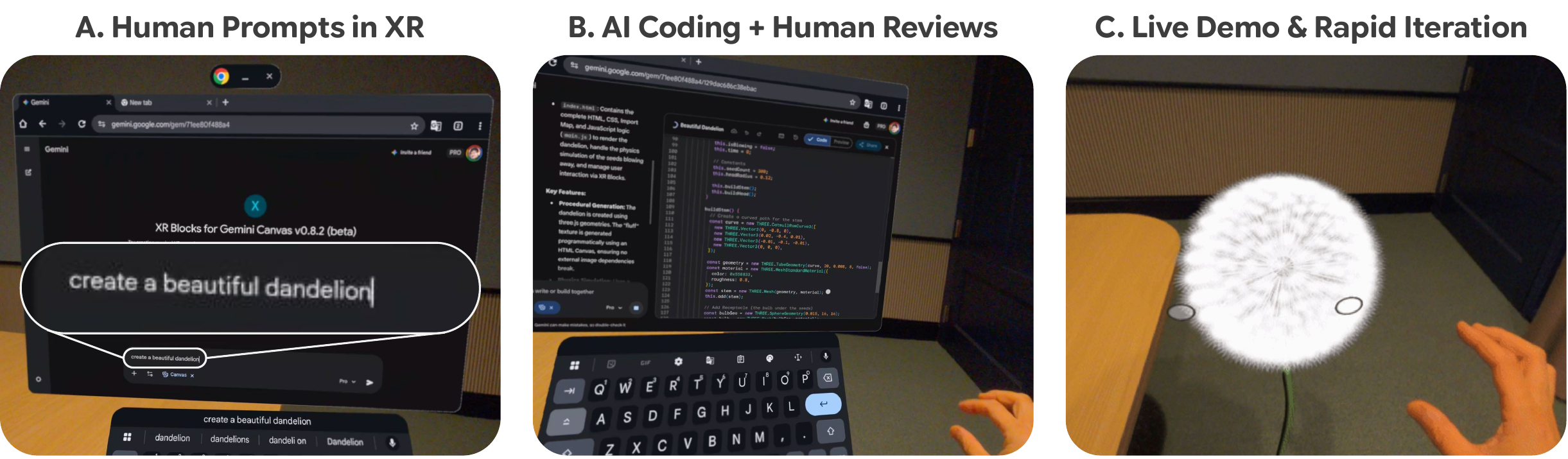}
  \caption{Example user journey of \thesystem, an end-to-end workflow for creating immersive AI + XR experiences via vibe coding: (A) User types ``create a beautiful dandelion'' with \xrblocks Gem (\gemurl) on a Galaxy XR headset in  Chrome. (B) Gemini translates the input into an interactive XR application within a minute, while the user reviews its reasoning and coding process. (C) User selects the ``Enter XR'' button and sees an animated dandelion dispersing upon pinch.}
  \Description[Three-panel figure of XR development workflow]{A three-panel figure illustrates an XR development workflow. Panel A, "Human Prompts in XR," shows a user's first-person view in an XR headset, typing the prompt "create a beautiful dandelion" into a floating virtual interface. Panel B, "AI Coding + Human Reviews," displays a virtual screen with code generated by an AI, which a human is reviewing. Panel C, "Live Demo & Rapid Iteration," presents the final result: a glowing virtual dandelion model placed on a physical floor, with a human hand reaching towards it.}
  \label{fig:teaser}
\end{teaserfigure}

\maketitle

\section{Introduction}

\begin{figure*}[t]
  \includegraphics[width=\textwidth]{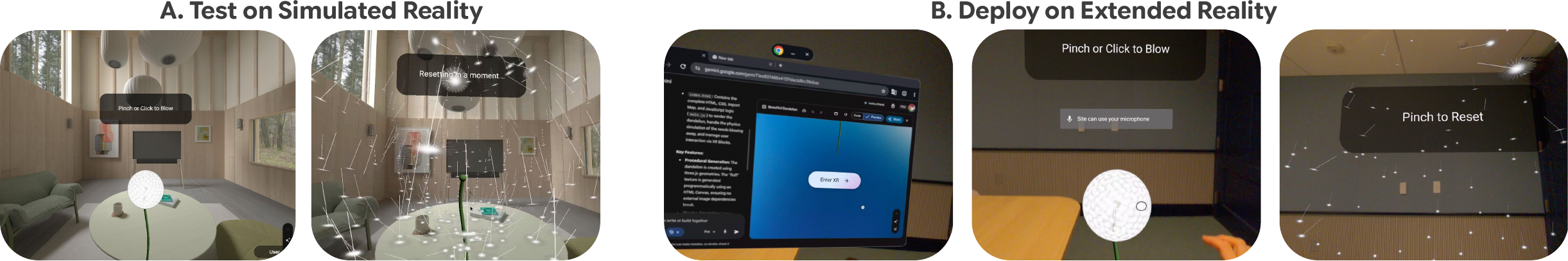}
  \caption{\thesystem accelerates AI + XR prototyping by allowing users to (A) test their ``vibe coding'' results on desktop in a ``simulated reality'' environment, and (B) deploy the same demo on an Android XR headset with body and hand interactions. }
  \Description[Comparison of simulated testing versus real-world deployment]{A two-part figure contrasts simulated testing with real-world deployment. Panel A, "Test on Simulated Reality," displays a virtual dandelion placed within a photorealistic 3D rendered living room, showing the flower both intact and with seeds dispersing. Panel B, "Deploy on Extended Reality," shows the application running in a physical room via an XR headset, featuring a code interface followed by the virtual dandelion overlaid on the real environment, where prompts like "Pinch or Click to Blow" guide the user to interact with the dispersing seeds.}
  \label{fig:intro}
\end{figure*}

Recent advances in Large Language Models (LLMs) \cite{Vaswani2017Attention,Brown2020Language,team2023gemini,Guo2025DeepSeek} and agentic workflows \cite{geminicli,antigravity,Cursor2025Cursor} are fundamentally reshaping software engineering and creative computing. We are witnessing the growing prevalence of ``vibe coding'' \cite{Edwards2025Will}, a paradigm in which high-level human intent is translated directly into functional software. While tools such as Gemini Canvas \cite{Google2025GeminiCanvas}, Antigravity \cite{antigravity}, Cursor \cite{Cursor2025Cursor}, and Claude Code \cite{claudecode} have successfully expanded this capability for 2D and, more recently, 3D web development~\cite{Vichare2025Webdev,Earle2025DreamGarden}, the domain of Extended Reality (XR) remains technically inaccessible for rapid prototyping. Creating \textit{intelligent} \textit{spatial} experiences still requires navigating a fragmented ecosystem of perception pipelines \cite{Zhu2025AgentAR,Shi2025Caring,He2023Ubi}, low-level sensor APIs \cite{Li2025EchoSight,Fang2020Wireality}, massive game engine hierarchies \cite{unity,unreal}, and the physical friction of constant on-device testing. 

Crucially, the absence of ``vibe coding'' in XR stems from a fundamental mismatch between how spatial software is currently built and how LLMs reason. Current models struggle to generate reliable XR applications because they are forced to operate over massive, low-level syntax trees and fragmented API layers rather than semantic, human-centered spatial concepts. This incidental complexity routinely exhausts context windows, fragments reasoning steps, and induces code hallucinations. To enable reliable intent-driven spatial computing, we need an abstraction layer explicitly designed not just for human accessibility, but for \textit{machine reasoning}. 

To bridge this gap, we introduce \xrblocks \citexrblocks
, an open-source WebXR~\cite{webxr} framework designed specifically to make spatial computing intelligible to generative AI. As the foundational contribution of this work, \xrblocks introduces an \textit{LLM-native} ``Reality Model'' that treats the user, the physical environment, and intelligent agents as first-class, configurable primitives. By encapsulating the complexities of sensor fusion, rendering, and spatial physics into a concise, semantic vocabulary, \xrblocks provides LLMs with the precise architectural constraints necessary to generate complex spatial behaviors without hallucinating underlying implementation.

Building upon this enabling framework, we present \thesystem, an end-to-end rapid prototyping workflow that leverages the long-context reasoning of LLMs to act as an expert XR designer. Using a specialized system prompt and a web-based interface, this workflow translates natural language directly into deployable, physics-aware XR applications in under 90 seconds. To solve the persistent HCI challenge of spatial testing friction, \thesystem allows creators to rapidly iterate on generated scripts within a desktop ``simulated reality'' environment before deploying instantly to an Android XR \cite{Google2025AndroidXR} headset for live hand interaction and environmental sensing.

Finally, evaluating the viability of generated XR software requires moving beyond subjective manual testing. We contribute the VCXR60 dataset, a pilot dataset consisting of 60 diverse spatial computing prompts sourced from four workshops. Coupled with an automated, headless-browser evaluation pipeline, we provide a quantitative baseline for XR code generation.

Our evaluation, alongside diverse application scenarios demonstrating XR realism and multi-modal interaction, illustrates the robustness of the system. 

In summary, we contribute:
\begin{itemize}
  \item \textbf{The \xrblocks Framework:} An open-source, LLM-native WebXR architecture that abstracts low-level spatial perception and interaction pipelines into a semantic ``Reality Model'' optimized for reliable AI code generation.
  \item \textbf{\thesystem Workflow:} A web-based, rapid prototyping loop that leverages LLMs to translate natural language intent into deployable XR applications, utilizing a novel desktop-simulator-to-headset testing pipeline to reduce iterative friction.
  \item \textbf{The VCXR60 Dataset \& Evaluation:} A preliminary dataset and automated testing pipeline for evaluating the runtime execution viability of intent-driven XR code generation, demonstrating the high one-shot pass rates of our framework.
\end{itemize}

As an initial exploration of vibe coding in XR, this work focuses on technical feasibility and automated runtime viability rather than formal user studies or cross-model evaluations. We hope our open-source framework and benchmark lay the groundwork for next-generation agentic workflows and reproducible AI + XR authoring.

\section{Related Work}
\thesystem bridges the gap between generative AI workflows and spatial computing by enabling intent-driven XR prototyping. 

\begin{table*}[!ht]
\centering
\scriptsize
\renewcommand{\arraystretch}{1.2} 
\setlength{\tabcolsep}{0.8em}     
\begin{tabular}{ l c c c c c c c c } 
 \toprule
 \textbf{System} & \begin{tabular}{@{}c@{}} \textbf{in-situ XR} \\ \textbf{prompting} \end{tabular} & \begin{tabular}{@{}c@{}} \textbf{web-native} \\ \textbf{architecture} \end{tabular} & \begin{tabular}{@{}c@{}} \textbf{LLM-optimized} \\ \textbf{abstraction} \end{tabular} & \begin{tabular}{@{}c@{}} \textbf{interactive} \\ \textbf{perception} \end{tabular} & \begin{tabular}{@{}c@{}} \textbf{intent-driven} \\ \textbf{vibe coding} \end{tabular} & \begin{tabular}{@{}c@{}} \textbf{desktop} \\ \textbf{simulator} \end{tabular} & \begin{tabular}{@{}c@{}} \textbf{zero-build} \\ \textbf{sharing} \end{tabular} & \begin{tabular}{@{}c@{}} \textbf{open} \\ \textbf{source} \end{tabular} \\ 
 \midrule
 Unity~\cite{unity} MCP & - & - & - & - & - & \Checkmark & - & - \\ 
 LLMR~\cite{DeLaTorre2024LLMR} & \Checkmark & - & - & limited & \Checkmark & - & - & \Checkmark \\
 DreamCodeVR~\cite{Giunchi2024Dreamcode} & \Checkmark & - & - & - & \Checkmark & - & - & \Checkmark \\
 agentAR~\cite{Zhu2025AgentAR} & \Checkmark & - & - & \Checkmark & \Checkmark & - & - & \Checkmark \\
 UniGen~\cite{yang2025unigen} & - & - & - & - & \Checkmark & \Checkmark & - & \Checkmark \\
 UnrealLLM~\cite{songtang2025unrealllm} & - & - & - & - & \Checkmark & - & - & \Checkmark \\
 AutoUE$^*$~\cite{yin2026autoue} & - & - & - & - & \Checkmark & \Checkmark  & - & \Checkmark \\
 A-Frame~\cite{AFrame} & - & \Checkmark & - & - & - & - & \Checkmark & \Checkmark \\
 Wonderland Engine~\cite{wonderland2026} & - & \Checkmark & - & - & - & - & - & - \\
 Meta IWSDK$^*$~\cite{meta2025iwsdk} & - & \Checkmark & ECS interop & - & \Checkmark & \Checkmark & - & \Checkmark \\
 \textbf{VCXR (XR Blocks)} & \Checkmark & \Checkmark & \Checkmark & \Checkmark & \Checkmark & \Checkmark & \Checkmark & \Checkmark \\
 \bottomrule
 \multicolumn{9}{l}{\scriptsize \textit{$^*$Denotes concurrent work to \xrblocks [9/2025] / \thesystem [2/2026] (e.g., Meta IWSDK [10/2025], AutoUE [4/2026]).}} \\
\end{tabular}
\caption{Comparison of key capabilities across XR authoring and generative AI systems. Evaluation criteria: (1) \textbf{in-situ XR prompting}: prompting directly within the XR headset session; (2) \textbf{web-native architecture}: execution entirely via WebXR without native binaries; (3) \textbf{LLM-optimized abstraction}: exposing a concise semantic syntax to minimize token boilerplate for XR interactions; (4) \textbf{interactive perception}: supporting real-time on-device models like gesture recognition and object recognition; (5) \textbf{intent-driven vibe coding}: end-to-end translation from natural language into interactive XR; (6) \textbf{desktop simulator}: built-in browser environment emulating head, hand, and sensor inputs; (7) \textbf{zero-build sharing}: distributable via instant URL without packaging; (8) \textbf{open source}: permissively licensed repository.}
\label{tab:vcxr-systems-comparison}
\end{table*}

\subsection{GenXR: Creating AI + XR Experiences}
XR development tooling has evolved from early toolkits like VR Juggler~\cite{Bierbaum2001VR} and ARToolKit~\cite{kato1999marker} to comprehensive game engines such as Unity~\cite{unity} and Unreal~\cite{unreal}. While these native engines offer high ceilings for fidelity, they impose steep learning curves, significant friction for rapid prototyping for XR, and high barriers for sharing executable code. Conversely, WebGL libraries like \texttt{three.js}~\cite{threejs} and A-Frame~\cite{AFrame} offer cross-platform accessibility, but lack the high-level abstractions necessary for complex XR interaction and seamless AI integration. 

Recent web-based solutions and SDKs have attempted to bridge this gap, including the Wonderland Engine~\cite{wonderland2026} and Meta's Immersive Web SDK (IWSDK)~\cite{meta2025iwsdk}. While Wonderland Engine is a powerful WebXR-compatible engine optimized for game making, it requires creators to download a localized desktop host and does not easily integrate with web-based AI workflows. IWSDK features a built-in desktop simulator and occlusion capabilities~\cite{Du2020DepthLab}, but crucially lacks the Interactive Perception (AI integration) required to treat the environment as a first-class primitive for agentic workflows; its lack of camera support also makes it challenging for AI and XR integration.

\xrblocks bridges this divide. By offering a high-level, web-optimized interaction model akin to MRTK~\cite{mrtk3}, VRTK~\cite{vrtk}, and XRI~\cite{xri} for Unity, our framework abstracts low-level sensor and perception pipelines so creators can focus on the \textit{what} of an experience rather than the \textit{how}. As a web-native architecture, \xrblocks makes AI + XR prototyping inherently accessible and reproducible, enabling the community to share, fork, and build upon each other's creations. \thesystem further leverages this framework to create behavioral AI + XR experiences, with complex 3D asset creation~\cite{Poole2022DreamFusion,tang2024lgm,xu2024grm,Hu2025Thing2Reality,hu2024generative,yu2023dynascape} remaining out of scope. While the concurrent work IWSDK also provides a high-level WebXR SDK, it usually requires a compilation step and an interop between \texttt{three.js}~\cite{threejs} and its own ECS subsystem, while lacking deeper perception runtimes such as gesture recognition and the native vision-language model integration offered by \xrblocks.

\subsection{Vibe Coding: Gaps from AI to XR}
The AI community benefits from a ``flywheel effect'' driven by open ecosystems such as Hugging Face~\cite{huggingface} and TensorFlow Hub~\cite{tensorflowhub}, crowd-sourced benchmarks like LMArena~\cite{Chiang2024Chatbot}, and composable frameworks including JAX~\cite{Jax2018Github}, PyTorch~\cite{Paszke2019PyTorch}, and TensorFlow~\cite{tensorflow2015-whitepaper}. XR, however, lacks a comparable substrate for rapid, intent-driven iteration on a reproducible platform.

Recent efforts have begun to address this by leveraging LLMs to generate spatial objects and scenes. Systems like LLMR~\cite{DeLaTorre2024LLMR}, DreamCodeVR~\cite{Giunchi2024Dreamcode}, and Thing2Reality~\cite{Hu2025Thing2Reality} target Unity, with commercial tools such as Bezi~\cite{BeziAIAssistance} pursuing a similar direction. On the infrastructure side, Ubiq-Genie~\cite{Numan2023UbiqGenie} proposed a server-client architecture that exposes AI services to Unity-based XR applications. Recent advancements also include AutoUE~\cite{yin2026autoue}, a multi-agent system that generates 3D games in Unreal Engine and utilizes an automated play-testing pipeline via a Model Context Protocol (MCP) to provide desktop simulation evaluation for the LLM. Furthermore, systems like DreamGarden~\cite{Earle2025DreamGarden} demonstrate LLM-driven hierarchical planning to generate functional 3D game environments and C++ logic in Unreal Engine.

While these approaches successfully push the boundaries of automated game design and 3D scene composition, they predominantly rely on asynchronous compilation workflows within heavy, closed-engine ecosystems like Unity and Unreal. As a result, they are frequently limited to one-time generation tasks and lack deep AI capabilities, such as generating an agent that can dynamically recognize objects in the physical room. This friction limits the collective ``vibe coding'' paradigm~\cite{Edwards2025Will}, in which an agent's utility is fundamentally tied to its ability to synthesize and rapidly share knowledge across the entire relevant domain.

To overcome this, we develop \xrblocks as a spatial computing platform on the open web inspired by Visual Blocks~\cite{Du2023Rapsai}. By coupling LLMs~\cite{Zhou2025InstructPipe} directly with our ``Reality Model'', \thesystem enables creators to generate immersive, intelligent, and physics-aware interactions. Unlike prior art, our framework supports in-situ XR prompting directly within the environment and features a seamless desktop simulator for frictionless testing with hand inputs, camera simulation, and cross-platform machine learning runtimes. This web-based approach ensures that every contribution is (1) built upon a transparent, web-standard codebase, (2) designed for rapid iteration via zero-build (contrasting Wonderland Engine and IWSDK) sharing, and (3) instantly distributable across diverse devices and users. By prioritizing these pillars, we move toward an ecosystem where the community does not merely consume agents but actively drives their evolution, supporting a ``flywheel effect'' for XR. Refer to \autoref{tab:vcxr-systems-comparison} for a recapitulation of key differentiators.

\begin{figure*}[t]
  \includegraphics[width=\textwidth]{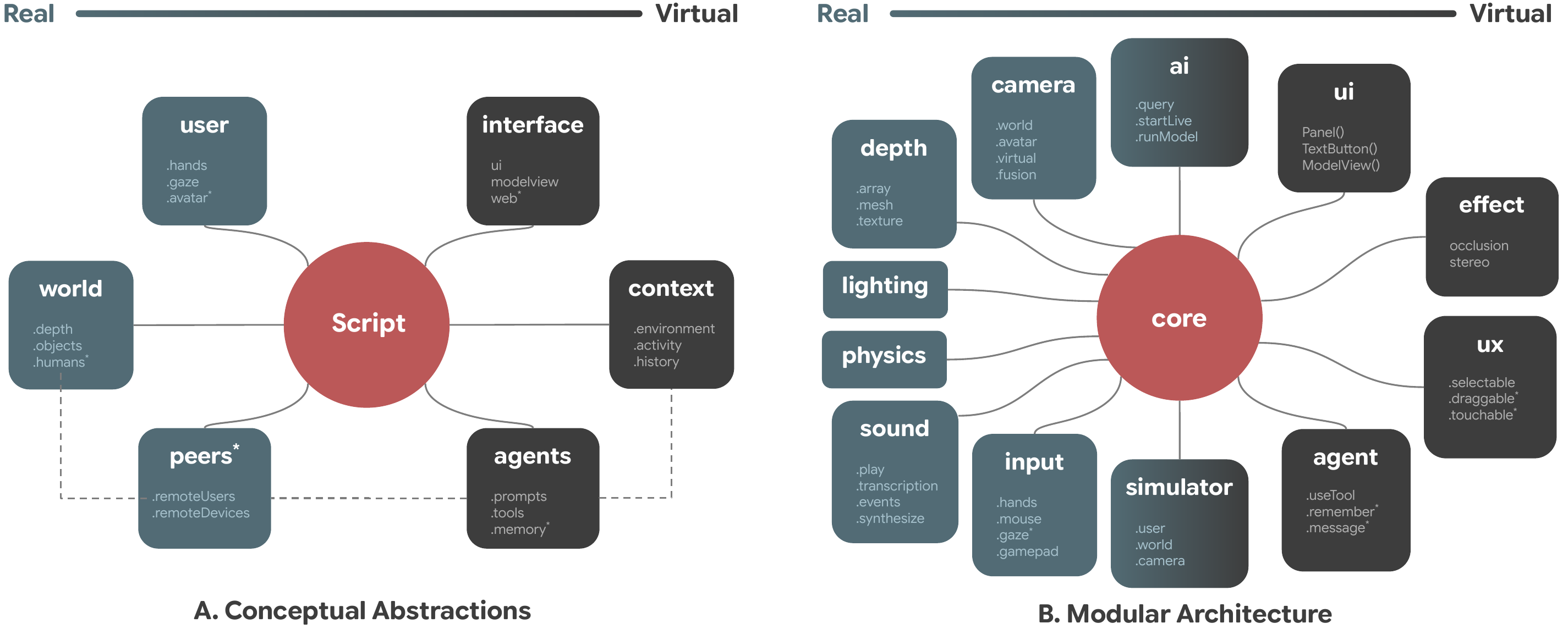}
  \Description[Diagram illustrating the conceptual and modular framework]{A two-part diagram illustrates the framework. Panel A, "Conceptual Abstractions," shows a central "Script" node connected to surrounding nodes representing "user", "world", "peers", "agents", "context", and "interface". Panel B, "Modular Architecture," features a central "core" node connected to surrounding technical modules.}
  \caption{Design of the \xrblocks Framework: (A) The ``Reality Model'' conceptual abstraction, which aligns spatial computing primitives 1:1 with natural language concepts to prevent LLM hallucination over fragmented syntax trees. (B) The modular architecture of the ``core'' engine encapsulating low-level perception and interaction logic. ${}^\ast$ denotes ``not yet implemented''.}
  \label{fig:framework_overview}
\end{figure*}

\section{System Architecture}
\label{sec:the_system}

We introduce \thesystem, a rapid prototyping workflow that combines the generative reasoning capabilities of LLMs with a high-level modular XR framework. This combination allows creators to bypass the friction of low-level systems programming and focus on designing spatial behaviors, AI logic, and interactions.

\subsection{The \xrblocks Reality Model}
At the foundation of our system is \xrblocks \cite{Li2025XR}, an open-source, LLM-native abstraction layer explicitly designed to make spatial computing intelligible to both human creators and generative AI. Traditional game engines and low-level WebGL APIs represent scenes as massive, deeply nested syntax trees and fragmented data streams from gamepad and hands. When forced to navigate these structures, LLMs routinely exhaust their context windows, lose track of spatial relationships, and hallucinate API calls. 

To solve this, \xrblocks introduces a semantic \textit{Reality Model}. This unified abstraction absorbs the incidental complexity of perception, rendering, and interaction behind high-level, human-centered primitives (\textit{e.g.}, \texttt{user}, \texttt{world}, \texttt{ui}, \texttt{ai}), providing a robust, concise vocabulary optimized for machine reasoning. By encapsulating WebXR session initialization, render loops, and input polling into declarative lifecycle hooks within \texttt{xb.Script}, the LLM spends its reasoning budget mostly on \textit{application logic} rather than \textit{boilerplate plumbing}. The following snippet demonstrates \xrblocks's design principle of keeping syntax semantically dense for the LLM. It creates an object at the user's eye level with a preferred distance, which changes color when the user performs a hand pinch. Crucially, this code is \textit{device-agnostic}---the \texttt{onSelectEnd} event fires seamlessly whether triggered by a physical hand gesture in an Android XR headset, a touch event on an Android phone, or a mouse click in the desktop simulator:

\begin{jscode}
class MainScript extends xb.Script {
  init() {
    // geometry and lighting setup is omitted.
    this.player = new THREE.Mesh(geometry, material);
    this.player.position.set(0, xb.user.height - 0.2, -xb.user.objectDistance);
    this.add(this.player);
  }
  onSelectEnd(event) {
    this.player.material.color.set(Math.random() * 0xffffff);
  }
}
\end{jscode}

As shown in \autoref{fig:framework_overview}, the framework's \textit{core} engine manages a rich, plug-and-play architecture of subsystems required for spatial computing, significantly expanding the bounds of what LLMs can instantly generate:

\begin{itemize}[leftmargin=*]
    \item \textbf{Hands \& Gestures:} Cross-platform 25-joint 6-DoF hand tracking, with real-time detection for gestures (\textit{e.g.}, pinch, grab, palm-up, point), alongside hooks for custom on-device models.
    \item \textbf{World Understanding \& Physics:} Real-time environmental depth meshing, dynamic occlusion, and surface plane estimation. Powered by the Rapier \cite{rapier} physics engine, virtual rigid bodies dynamically collide with and bounce off physical room geometry.
    \item \textbf{Spatial UI \& Audio:} Declarative 3D flexbox panels with Signed Distance Field (SDF) rendering for crisp text, combined with speech synthesis, and voice recognition APIs.
    \item \textbf{AI Integration:} A hybrid edge-cloud architecture supporting both on-device models in LiteRT.js and cloud multimodal LLMs (Gemini) for live spatial reasoning and vision-language queries.
    \item \textbf{Multiplayer (Peers):} Built-in WebRTC networking for cross-platform presence, shared spatial state objects, and spatialized voice to facilitate multi-user collaborative experiences.
\end{itemize}

By abstracting these technical hurdles, \xrblocks lowers the barrier to entry while retaining a high ceiling for expressivity, positioning it as the ideal foundation for \thesystem. The framework's architecture was iteratively refined through the development of over 30 open-source templates (examples: \autoref{fig:samples}), incorporating feedback from weekly workshops and live demos. 

\subsection{Agentic Alignment \& System Prompt}

\thesystem leverages the long-context capabilities and advanced reasoning processes of modern LLMs like Gemini~\cite{team2023gemini} to function as an expert XR designer and engineer. However, merely providing an API is insufficient; the LLM must be explicitly constrained and guided to ensure reliable code generation. 

To achieve this, we first developed a specialized system prompt architecture (detailed in \autoref{sec:systemprompts}) that explicitly grounds the LLM in the \xrblocks Reality Model through an open-source web interface. As the framework expanded, we formalized these prompt-driven constraints natively into the repository as \textbf{Agent Alignment Specifications} (\textit{e.g.}, \texttt{AGENTS.md}). These machine-readable files now serve as the ultimate source of truth for downstream AI coding agents, automatically teaching them the repository structure, key files, testing protocols, and spatial conventions. The foundational system prompt architecture consists of:

\begin{figure}[t]
  \includegraphics[width=\linewidth]{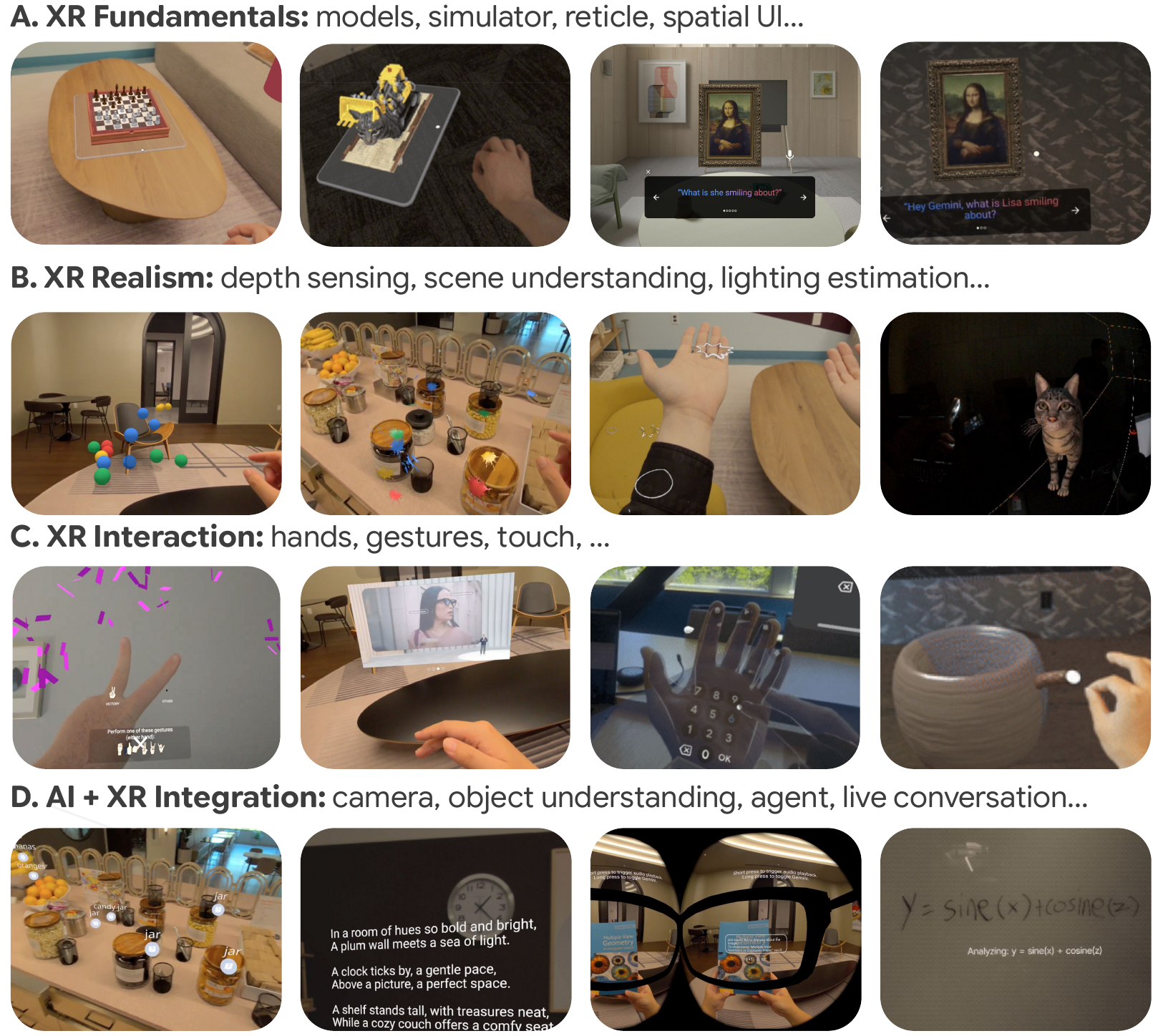}
  \Description[Grid of screenshots showcasing XR capabilities]{A grid of example screenshots showcases XR capabilities across four categories: (A) XR Fundamentals, (B) XR Realism, (C) XR Interaction, and (D) AI + XR Integration.}
  \caption{Human-coded templates and samples in \xrblocks provide the foundational best practices and API grounding for \thesystem.}
  \label{fig:samples}
\end{figure}

\begin{itemize}[leftmargin=*]
    \item \textbf{Persona \& Guidelines:} Configures the LLM with a domain expert persona adhering to the rules of engagement in the system prompt, focusing on best practices for room-scale XR environments (\textit{e.g.}, spatial layout, human-scale proportions, safety bounding, and interaction distances).
    \item \textbf{Architectural Constraints:} Dictates how dependencies and subsystems (\textit{e.g.}, \texttt{xb-core}, \texttt{xb-ui}, \texttt{xb-ai}) are handled to enforce the repository's strict declarative styles, directly informed by the system prompt to ensure generated code executes flawlessly without context loss.
    \item \textbf{Source Code \& Templates:} Grounds the model with the full source code of the core \texttt{xrblocks.js} library, alongside curated templates and samples (illustrated in \autoref{fig:samples}). This strict grounding acts as a technical constraint---enforcing the ``only call APIs that exist'' rule to minimize hallucination and encourage adherence to established spatial design patterns.
\end{itemize}

\subsection{The Simulated-to-Extended Reality Loop}
Users interact with \thesystem through high-level natural language prompts (or ``vibes''). The LLM translates this intent directly into functional \xrblocks scripts~\cite{Zhou2025InstructPipe}. Crucially, to solve the persistent HCI friction of iterative spatial testing---where creators must constantly compile, build, and don/doff a headset---\thesystem introduces a seamless \textit{simulated-to-extended reality loop}. 

Scripts can be previewed immediately in a built-in desktop simulated reality environment running directly in the browser. This
injects a proxy of the physical environment (complete with mock
depth maps and scene meshes). This allows for rapid visual validation of physics, logic, spatial UI layout, and AI reasoning without leaving the desktop. Once verified, the exact same URL is deployed instantly to an Android XR headset (\textit{e.g.}, Galaxy XR) for real-world environmental sensing and live hand interaction. This cross-platform, write-once-run-anywhere pipeline eliminates build times and enables a frictionless ``vibe coding'' loop.

\subsection{Application Scenarios: From Prompt to XR}
\label{sec:application_scenarios}
\label{sec:workshop_feedback}
To demonstrate the versatility and expressive range of \thesystem, we present representative application prototypes generated during four hands-on workshops with 20 participants. These prototypes (with full prompts listed in \autoref{sec:application_scnarios_prompts}) exercise the core capabilities of the \xrblocks Reality Model across interactive learning, room-scale physics, and procedural gaming (\autoref{fig:examples}):

\begin{figure}[t]
  \includegraphics[width=\linewidth]{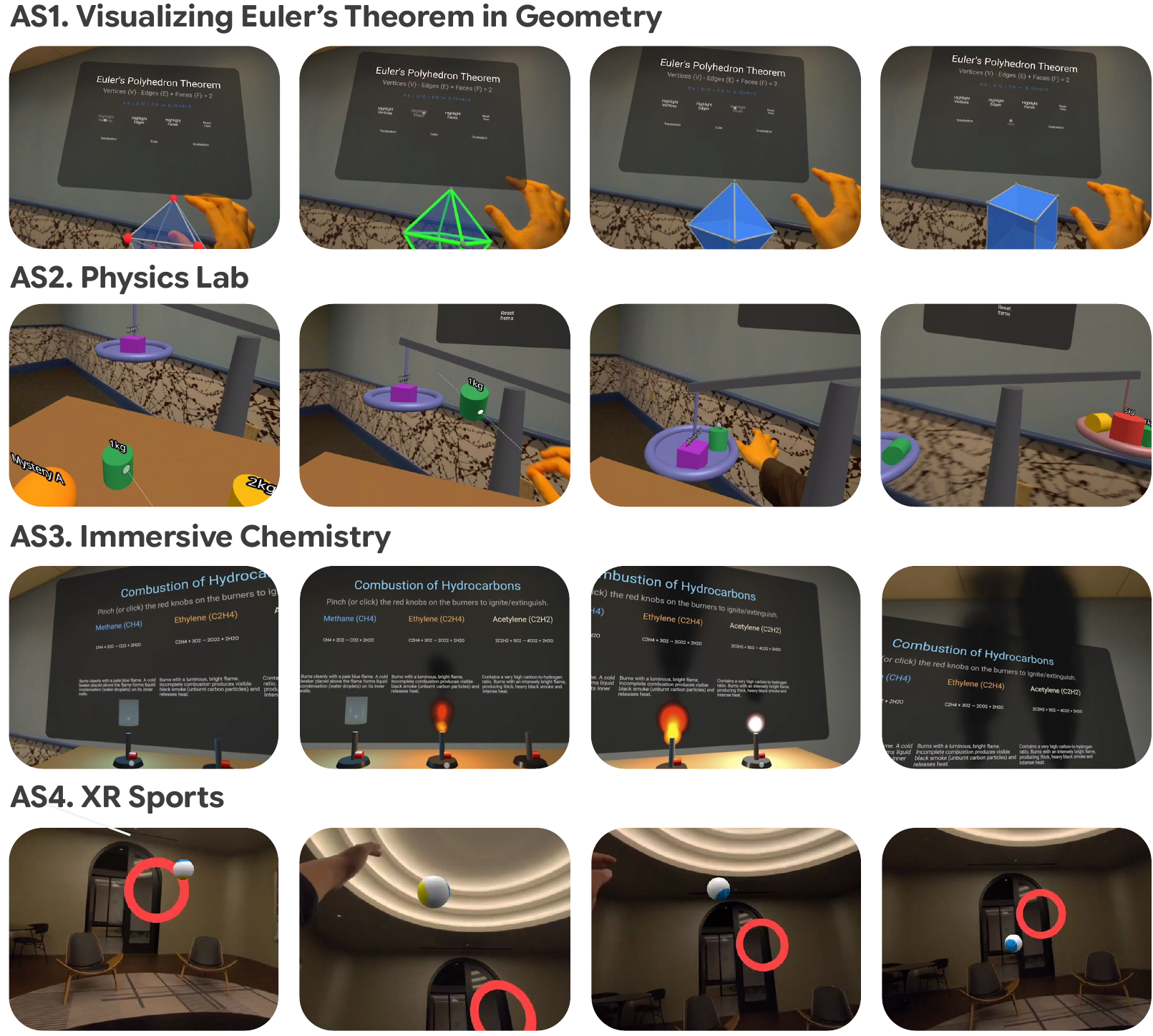}
  \Description[Grid of screenshots of generated applications]{The image is a grid of sixteen screenshots divided into four horizontal rows showing first-person perspectives of educational and sports applications.}
  \caption{AI-generated application for educational and fitness use cases. See Appendix B for full prompts.}
  \label{fig:examples}
\end{figure}

\paragraph{\textbf{Educational AI + XR Experiences}}
As shown in \autoref{fig:examples}, \thesystem supports rapid prototyping of interactive spatial learning environments. Prompted with \textit{``visualize Euler's theorem in geometry and explain vertices, edges, and facets''} (AS1), the system generated an immersive \textit{Math Tutor}. The LLM instantiated polyhedral geometries (tetrahedron, cube, octahedron), wired pinch interactions via \texttt{onSelect()} to toggle face and edge highlights, and rendered accompanying spatial explanation cards using \texttt{xb.ui}. Similarly, prompting for an \textit{Immersive Chemistry} experience (AS3) produced educational formula cards and 3D volumetric particle effects, enabling users to trigger and observe simulated hydrocarbon combustion (e.g., methane) safely via hand-pinch gestures.

\paragraph{\textbf{Physics-Aware XR Realism \& Interaction}}
By grounding the model in \xrblocks's physics and depth subsystems, users can generate tactile, environment-aware experiences (\autoref{fig:examples}). Prompted for an \textit{XR Sports} application (\textit{``Let me play volleyball with hands and collide with my environment''}, AS4), the system generated a textured volleyball that reacts dynamically to both 6-DoF tracked hand joints and real-world room geometry extracted via depth sensing (\texttt{xb.depth.mesh}). Furthermore, a \textit{Physics Lab} prompt (AS2) generated a mixed-reality balance scale where users grasp, lift, and drop labeled weights to explore mechanical equilibrium in real time.

\paragraph{\textbf{Game Prototyping \& Procedural Generation}}
The workflow drastically compresses the time required to stand up interactive XR game mechanics. Prompted to \textit{``create the Chrome Dino game in XR''} (AS5), \thesystem generated a voxelized runner complete with rushing obstacles on a floating translucent track, spatial audio cues, and collision triggers in under 90 seconds. Prototypers also leveraged the workflow for procedural spatial compositions, including rendering interactive astronomical star charts and explorable architectural structures directly in the headset.

\section{Preliminary Evaluation with VCXR60}
\label{sec:evaluation}

Evaluating XR applications poses unique challenges, as it typically requires manual, on-device testing and subjective human evaluation. However, assessing the viability of generative XR systems requires fundamentally new, scalable technical metrics that can keep pace with the rapid iteration of large language models. To meet this need, we introduce the \textbf{VCXR60} dataset, an open-source benchmark designed to automate the testing of intent-driven spatial computing.

Sourced from four one-hour workshops, VCXR60 consists of 60 XR application prompts provided by 20 participants (detailed in \autoref{sec:vcxr60_dataset}). Using this dataset with \xrblocks's desktop simulator, we measured inference time and one-shot pass rates (pass@1).
Inspired by code LLM benchmarks like HumanEval~\cite{chen2021evaluatinglargelanguagemodels}, we define a ``pass'' strictly as a zero-error execution in an automated runtime test. In 3D graphics and WebXR, the rendering pipeline serves as an initial syntactic and execution validator, as runtime errors during initialization, asset loading, or per-frame rendering immediately flag code failures. We monitor uncaught exceptions and console error logs across scene initialization and active animation frames in a headless Chromium browser via Playwright~\cite{Playwright}. Crucially, we emphasize that the pass@1 metric is solely an automated runtime check (i.e., verifying that the generated code compiles, initializes, and runs without uncaught errors or crashing); it is not a measure of semantic correctness, spatial fidelity, or user intent fulfillment.

Early exploratory testing in September 2025 using \xrblocks v0.1.0 with Gemini 2.5 Pro yielded an approximate 70\% pass rate, where failures predominantly stemmed from edge-case framework bugs (e.g., coordinate frame mismatches in the simulator and missing gesture callbacks) and LLM API hallucinations. These insights informed a rapid six-month iteration cycle. After 10 major releases, we evaluated \xrblocks v0.11.0 on VCXR60 using the latest available models at the time of benchmarking: \texttt{gemini-3.1-pro-preview} and \texttt{gemini-3-flash-preview} (accessed March 19--22, 2026). We selected these preview backends because their extended reasoning budgets (thinking mode) were essential for complex multi-component spatial scene synthesis. While preview endpoints undergo deprecation cycles as stable models emerge, we ensure full empirical reproducibility by releasing the versioned SDK and the complete prompt suite on our project repository. We also note a methodological caveat: the improvement from $\sim$70\% in v0.1.0 to $>$87--95\% in v0.11.0 reflects the combined benefit of framework-level hardening (e.g., standardized schemas, curated system instructions, and simulator bug fixes) and general advancements in foundation model capabilities over that six-month period.

Averaged across five runs, our evaluation revealed a clear performance advantage for larger models, while differences between ``thinking'' levels within the same model family were minor and statistically non-significant (\autoref{tab:sys_results}). Specifically, Gemini 3.1 Pro achieved higher overall runtime pass rates (High: $95.5\% \pm 2.2\%$, 95\% CI [$92.7\%, 98.3\%$]; Low: $94.0\% \pm 2.6\%$, 95\% CI [$90.7\%, 97.3\%$]) compared to Gemini 3 Flash (High: $87.8\% \pm 3.4\%$, 95\% CI [$83.6\%, 91.9\%$]; Low: $87.3\% \pm 6.4\%$, 95\% CI [$79.4\%, 95.3\%$]). Because the confidence intervals for High and Low thinking heavily overlap, thinking mode does not significantly alter zero-error runtime pass rates; rather, its primary trade-off is inference latency versus planning capacity for complex prompts. Notably, across all backends, every single test prompt achieved at least one successful execution across the five runs (100\% at-least-one pass rate), indicating that multi-turn retry loops can effectively overcome non-deterministic generation errors. For simpler interactions (e.g., \textit{``Create a beautiful dandelion that blows away when I pick it up''}), faster models like Gemini Flash often completed generation in under 20 seconds. For applications requiring complex animations and precise hand-interaction logic (e.g., Prompt 060 in \autoref{sec:vcxr60_dataset}), however, Flash exhibited higher error rates compared to Gemini Pro.

\begin{table}[t]
\centering
\caption{Inference time and one-shot pass rate (pass@1) for XR Blocks Gem on VCXR60 across 5 runs (accessed March 19--22, 2026). Pass@1 reports the mean, standard deviation (SD), and 95\% confidence interval (CI) across runs. Note that pass@1 is solely a runtime check (i.e., code executes without crashing or unhandled console errors), not a measure of semantic correctness or alignment with user intent.}
\label{tab:sys_results}
\vspace{0.3em}
\scalebox{0.81}{
\begin{tabular}{llcccc}
\toprule
\multicolumn{2}{c}{\textbf{\thesystem}} & \multicolumn{2}{c}{\textbf{Time} (secs)} & \multicolumn{2}{c}{\textbf{Pass Rate@1} (5 runs)} \\
\cmidrule(lr){1-2} \cmidrule(lr){3-4} \cmidrule(lr){5-6}
Model Backend & Thinking & Median & IQR & Mean $\pm$ SD & 95\% CI \\ 
\midrule 
gemini-3.1-pro & High & 86.02          & 32.12         & \textbf{95.5\%} $\pm$ 2.2\% & [92.7\%, 98.3\%] \\
gemini-3.1-pro & Low  & 33.39          & 8.60          & 94.0\% $\pm$ 2.6\%          & [90.7\%, 97.3\%] \\ 
gemini-3-flash & High & 22.26          & 4.66          & 87.8\% $\pm$ 3.4\%          & [83.6\%, 91.9\%] \\ 
gemini-3-flash & Low  & \textbf{17.30} & 4.00          & 87.3\% $\pm$ 6.4\%          & [79.4\%, 95.3\%] \\ 
\bottomrule 
\end{tabular}
}
\end{table}

\paragraph{\textbf{Manual Review and Qualitative Observations}}
To evaluate interaction quality beyond automated headless checks, the authors manually reviewed and interacted with all generated demos across the 60 benchmark prompts in both the desktop simulator and on-device with an Android XR headset, selecting representative prototypes to showcase in the supplementary video. While all prompts achieved technical runtime viability in at least one run, not every creation was qualitatively successful due to non-determinism in LLM inference. For instance, in a two-player ``ping-pong game'' (Prompt 055 in \autoref{sec:vcxr60_dataset}), the generated racket was occasionally not grabbable by the user's hands; even when grasping succeeded, the racket sometimes failed to orient correctly with the hand joints, or the ping-pong ball failed to register physical collisions. These subtle spatial and behavioral discrepancies demonstrate that technical zero-error execution does not guarantee full interaction fidelity or user intent fulfillment, motivating future work on more rigorous spatial benchmarking platforms that systematically evaluate and align with human ratings (\autoref{sec:discussion_benchmarking}).

\paragraph{\textbf{Qualitative Feedback from Workshop Prototypers}}
\label{sec:workshop_feedback}
To complement our technical benchmarking with user perspective, we documented qualitative observations across four prototyping workshops involving 20 participants (spanning research scientists, UX prototypers, and software engineers). Three consistent themes emerged:
(1)~\textbf{Rapid Scaffolding vs. Native Build Latency}: Participants highlighted that generating functional 3D scenes in under 90 seconds dramatically bypassed the traditional 5--10 minute compile-and-deploy cycles of native engines like Unity, making rapid concept exploration viable.
(2)~\textbf{Friction in Fine-Grained Spatial Tuning}: While coarse scaffolding was rapid, participants reported friction when making localized adjustments (e.g., nudging coordinates or styling UI cards), noting that conversational prompting often unintentionally perturbed existing scene elements and motivating hybrid direct-manipulation controls.
(3)~\textbf{Debugging Interaction and Session Drift}: Participants found debugging dynamic spatial behaviors (e.g., hand triggers and colliders) challenging because the LLM lacks runtime visual perception, emphasizing the need for multimodal simulator snapshots and MVC architectures to decouple application state from transient LLM context.

\section{Deployment Case Studies and Reflections}

Following its open-source release, \xrblocks was adopted by early practitioners and workshop developers, yielding emergent prototyping patterns that bridge declarative 2D tooling with spatial physics. We summarize three qualitative deployment scenarios below:

\begin{figure}[h]
  \includegraphics[width=\linewidth]{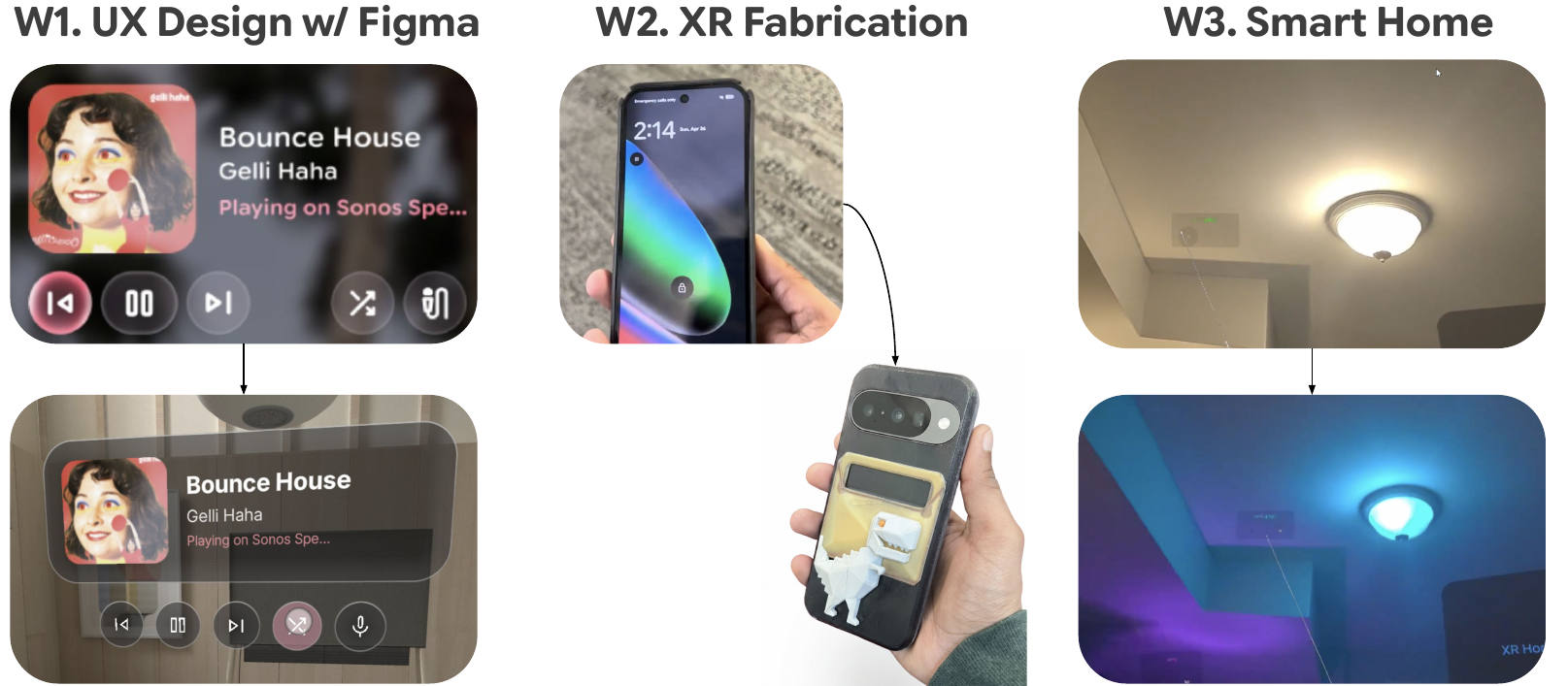}
  \Description{A three-part composite image illustrating real-world applications of the system. The left panel, labeled 'W1. UX Design w/ Figma', shows a 2D music player interface with play controls transforming into a floating, transparent 3D spatial UI within a physical room. The middle panel, 'W2. XR Fabrication', shows a 3D model of a phone case overlaid onto a real smartphone. The right panel, 'W3. Smart Home', displays an augmented reality interface used to control a physical environment.}
  \caption{Emergent in-the-wild applications of \thesystem. (W1) Translating a flat 2D UI mockup (top) into an interactive, translucent spatial UI. (W2) Validating the scale and fit of a generative 3D-printed phone case against a physical device measured in XR with depth sensing. (W3) Opportunistic smart home control linking spatial interactions with physical IoT devices.}
  \label{fig:wild}
\end{figure}

\paragraph{\textbf{Vibe Coding XR for Designers}}
Bridging the gap between declarative 2D design and spatial interaction has historically been a significant bottleneck. User experience designers typically author interfaces in 2D tools (\textit{e.g.}, Figma), leaving XR prototypers to manually recreate these layouts within Unity to achieve spatial interactivity---a process that often takes hours. Leveraging the latest coding agents alongside \xrblocks's UI modules, creators can now upload a flat mockup and simply prompt the system to ``create a spatial UI with the UI blocks skills.'' As demonstrated in \autoref{fig:wild} (W1), \thesystem ingests a 2D screenshot of a music player and translates it into a translucent, floating spatial interface, complete with physics-aware, pinch-interactive buttons.

\paragraph{\textbf{Vibe Coding XR for Fabrication}}
\thesystem also enables creators to transition virtual artifacts into the physical world. In one instance, an author prompted the system for a ``Chrome Dino phone case with a leather-style wallet for a Pixel 10.'' Gemini generated a reference image, which was subsequently lifted into a 3D mesh utilizing the TRELLIS~\cite{xiang2024trellis} generation model. Crucially, rather than judging proportions on a flat desktop monitor, the creator utilized the Android XR headset's passthrough to superimpose the generated 3D model directly onto a physical Pixel phone (\autoref{fig:wild}, W2). This \textit{in-situ} registration allowed the creator to rigorously validate the shape, true physical scale, and wallet fitment prior to exporting the asset for 3D printing. By sharing a common spatial representation, \thesystem minimizes the costly trial-and-error typically associated with digital fabrication.

\paragraph{\textbf{Vibe Coding XR for Home Controls}}
Extending spatial computing into the Internet of Things (IoT), an expert user deployed \xrblocks to interface with their Matter-enabled smart home ecosystem (\autoref{fig:wild}, W3). Using \thesystem, the user leveraged video see-through capabilities to authenticate the XR application's access to their local IoT network via a mobile companion device. By integrating real-time object recognition, the application identifies physical lighting fixtures within the room, assigns them to the existing smart home setup, and allows the user to manipulate colors and intensities through direct, \textit{in-situ} XR interactions. This compellingly demonstrates \thesystem's capacity to facilitate opportunistic, context-aware control of the physical environment.

\paragraph{\textbf{Engineering Reflections and Lessons Learned}}
Our design and deployment strategy for \xrblocks was inspired by open-web systems like Python Tutor~\cite{Guo2021Ten}. However, navigating the intersection of generative AI and WebXR revealed non-obvious engineering pitfalls and design pivots:

\textbf{The Cost of Monolithic Abstractions:} Early framework iterations attempted to wrap Three.js in a monolithic engine class that managed all lifecycles internally. In practice, this bloated context windows and caused severe API hallucinations. We pivoted to a lightweight \texttt{xb.Script} subclassing architecture where developers and models import only the specific primitives needed, drastically lowering token overhead.

\textbf{The Simulation-to-Device Fidelity Gap:} We initially assumed 2D mouse clicks would suffice as a testing proxy for hand gestures on desktop. However, upon deploying to physical headsets, subtle spatial bugs---such as inverted hand-joint orientation matrices, missing collision volumes, and incorrect depth occlusion---were immediately exposed. This necessitated re-architecting the desktop simulator to inject full 6-DoF simulated hand skeletons, synthetic depth maps, and camera streams into the WebXR polyfill.

\textbf{Learning from Recurring Hallucinations:} In generative code synthesis, LLMs frequently invent non-existent method signatures based on statistical priors from mature libraries (e.g., hallucinating \texttt{xb.hands.getPinchDistance()} or \texttt{xb.ui.addSlider()}). Rather than treating these solely as errors to be suppressed, we observed that when multiple models consistently hallucinated the exact same API shape across diverse prompts, it served as an unsupervised signal of an ergonomic API gap. We systematically cataloged these recurrent, syntactically coherent hallucinations and implemented them as first-class methods in subsequent framework releases.

\textbf{Balancing Web Standards with Modern Tooling:} While Guo~\cite{Guo2021Ten} advocates for older, conservative technologies to maximize longevity, generative AI advances at a pace that demands strict type safety. We adopted TypeScript internally to generate formal declaration files, which proved essential for grounding coding agents, while preserving a zero-build vanilla JavaScript bundle for end-user HTML deployments.

\section{Limitations and Discussion}

This paper presents our initial step towards bridging the ``idea to reality'' gap in spatial computing. While our current implementation successfully prioritizes core abstractions for generative XR, its limitations highlight critical avenues for future research.

\paragraph{\textbf{Absence of User Studies}}
While our weekly workshops provided rich qualitative feedback from 20 practitioners, this paper does not present a formal, controlled behavioral user study. Evaluating intent-driven spatial computing requires rigorous investigation into usability (e.g., System Usability Scale), learnability curves for non-programmers, cognitive workload (NASA-TLX), and task completion times relative to traditional game engines. Although technical systems research often focuses initially on architectural feasibility and reproducible benchmark metrics~\cite{fogarty2017code,hudson2014concepts}, human-in-the-loop usability remains an essential requirement for validating end-user authoring workflows.

\paragraph{\textbf{Narrow Scope of Automated Technical Evaluation}}
Our automated benchmark on VCXR60 captures baseline code viability via headless execution logs, verifying that generated applications compile, initialize, and run without unhandled runtime exceptions. However, this metric does not evaluate visual aesthetics, render-loop performance, physical plausibility, or semantic adherence to user intent. As demonstrated in our manual review, applets can achieve zero-error compilation while failing critical interaction heuristics (e.g., ungrabbable ping-pong rackets or misaligned joint transforms). Future benchmarks must incorporate multimodal autoraters and automated spatial unit tests.

\paragraph{\textbf{Dependence on a Single Model Family}}
All technical evaluations in this work were conducted using the Gemini model family. While these models demonstrate high one-shot viability when grounded in our framework, instruction following, spatial math competence, and API adherence may vary significantly across other frontier foundation models. Evaluating \xrblocks across diverse LLMs is a critical next step to verify framework generality.

\paragraph{\textbf{Performance and Platform Trade-offs}}
Our reliance on web technologies intentionally prioritizes accessibility and cross-platform shareability, though this introduces known trade-offs. The WebXR framework currently cannot match the raw rendering throughput of native engines like Unity or Unreal, and the dependency on cloud-hosted LLMs introduces network latency during the generation phase. To bridge this gap, a compelling longer-term objective is the development of an \textit{LLM-driven cross-compiler} capable of translating high-level \xrblocks scripts into optimized, native code for target engines such as Godot~\cite{godot}. We actively invite the community to contribute to the SDK to establish shared conventions for AI + XR interactions, which will also serve to optimize prompt efficiency and reduce LLM token costs.

\paragraph{\textbf{Toward Rigorous Benchmarking}}
\label{sec:discussion_benchmarking}
The VCXR60 dataset covers a valuable but limited slice of the vast AI + XR design space. Furthermore, our automated pass/fail metric captures baseline code viability but cannot assess usability, aesthetic quality, or interaction fidelity. As observed during our manual inspection of generated applets, subtle spatial failure modes---such as improper hand attachment, raycasting misalignments, or physics collision omissions---frequently escape automated runtime validation. A more rigorous future evaluation should decouple the contributions of the \xrblocks abstraction layer from the underlying \texttt{three.js} library, quantifying the exact token savings and error reduction our framework provides relative to raw API usage. Additionally, future benchmarks must address iterative, multi-turn usage scenarios---such as refining an existing prototype through conversational prompting---which introduces distinct challenges in context management and incremental code coherence. Just as ImageNet~\cite{Deng2009ImageNet} galvanized progress in computer vision, the XR and HCI fields urgently need large-scale, open-source datasets of \textit{prompt-to-interaction} pairs alongside automated spatial autoraters calibrated to human quality judgments to enable the systematic evaluation of generated spatial behaviors.

\paragraph{\textbf{Expanding the Interaction Vocabulary}}
Currently, \thesystem accepts text as its primary input modality. However, spatial intent is often difficult to articulate through text alone. Hybrid approaches that blend natural language with spatial UI selections, gaze targeting \cite{lee2024gazepointar,ramiotis2025context}, micro-gestures \cite{haneling2025beyond}, and cross-device interactions \cite{Zhu2025Beyond} could substantially improve the precision of user intent. Furthermore, \xrblocks is designed to be extensible beyond audio-visual outputs; integrating modules for haptics, olfactory interfaces, or other sensory modalities (e.g., via Arduino hardware through WebUSB) would empower creators to compose significantly richer, multi-sensory experiences.

\paragraph{\textbf{Accessibility and Responsible Deployment}}
As we evolve \xrblocks through open-source collaboration, we recognize that the framework currently lacks ready-to-use accessibility modules for motor or visual impairments---a critical gap given our goal of lowering technical development barriers for XR. Moreover, the responsible deployment of LLM-driven spatial authoring tools demands rigorous human-in-the-loop evaluation to understand how designers and novices co-create with AI agents. Ensuring that generative outputs safely augment human intent, rather than introducing harmful or disorienting spatial behaviors, remains a paramount challenge for the HCI, XR, and AI communities.

\section{Conclusion}

We presented \thesystem, an end-to-end workflow that translates high-level creative intent into functional XR prototypes. By coupling the generative reasoning of LLMs with \xrblocks, an LLM-native abstraction layer, we demonstrated an approach that substantially reduces the distance between a natural language concept and an interactive, physics-aware spatial application. Our preliminary evaluation on the VCXR60 dataset demonstrates that providing LLMs with a semantic ``Reality Model'' and concise, declarative component lifecycles yields high zero-error runtime execution pass rates (up to 95.5\%) across diverse spatial interaction scenarios, bypassing \textit{the compilation overhead} and \textit{steep learning curves} of traditional native game engines. Furthermore, by integrating a simulated-to-extended reality loop, we mitigate the substantial friction of iterative on-device testing.

We release this framework, workflow, and benchmark dataset as an invitation. We welcome the HCI, XR, and AI communities to build upon this foundation, expand the capabilities of generative spatial computing, and collectively work toward a future where moving from ``idea to reality'' is as fluid as describing it.

\ifanonymous
\else
\begin{acks}
We thank all of XR Blocks contributors in 2025: David Li and Ruofei Du (equal primary contributions), Nels Numan, Xun Qian, Yanhe Chen, and Zhongyi Zhou, (equal secondary contributions, sorted alphabetically), as well as Evgenii Alekseev, Geonsun Lee, Alex Cooper, Brandon Jones, Min Xia, Scott Chung, Jeremy Nelson, Xiuxiu Yuan, Jolica Dias, Tim Bettridge, Benjamin Hersh, Michelle Huynh, Konrad Piascik, Ricardo Cabello, and David Kim. We extended our thanks to all of our 2026 student researchers and GitHub contributors: William Huang, Xincheng Huang, Hyunsung Cho, Salman Chishti, Akilesh, Jamie; We further thank the Gemini Canvas and AI Studio teams for their support including, but not limited to: Tim Bettridge, Yan Li, Daniel Marques, Deven Tokuno, Levent Yilmaz, Saravana Rathinam, Samuel Petit, Mike Taylor-Cai, Ammaar Reshi, and Robert Berry, We would like to thank Mahdi Tayarani, Max Dzitsiuk, Jim Ratcliffe, Patrick Hackett, Seeyam Qiu, Coco Fatus, Alon Hetzroni, Aaron Kim, Yinghua Yang, Brian Collins, Eric Gonzalez, Keith Moon, Nicolás Peña Moreno, Yidang Zhang, Yuhao He, Yi-Fei Li, Ziyi Liu, Jing Jin for their feedback and discussion on our early-stage proposal and WebXR experiments, as well as all of DepthLab~\cite{Du2020DepthLab}, Ad hoc UI~\cite{Du2022Opportunistic}, Rapsai (Visual Blocks)~\cite{Du2023Rapsai}, InstructPipe~\cite{Zhou2025InstructPipe}, DialogueLab~\cite{Hu2025DialogLab}, AgentHands~\cite{Liu2026AgentHands}, and Sensible Agent~\cite{Lee2025Sensible} co-authors, which greatly inspired this project along the way. We appreciate all anonymous reviewers', Tim Herrmann and Andrew Helton’s thoughtful reviews. We thank Maryam Sanglaji, Max Spear, Adarsh Kowdle, and Guru Somadder, Shahram Izadi for the directional feedback and contribution.
\end{acks}
\fi

\bibliographystyle{ACM-Reference-Format}
\bibliography{99_vcxr}

\newpage

\appendix

\section*{Appendix for Vibe Coding XR}

\section{VCXR60 Dataset Prompts}
\label{sec:vcxr60_dataset}

\begin{description}
  \item[001: blowing\_dandelion] Create a beautiful dandelion. when I pick it up and hold with my hands, make the seeds blow away.
  \item[002: rainbow\_pen] Make a pen that draws rainbows in 3D.
  \item[003: vine\_growth] Create a vine growth animation with tendrill curling, leaf, sprouting, wall climbing, flower blooming, and organic spreading pattern.
  \item[004: bubble\_pop] Make a bunch of bubbles that pop when I touch them.
  \item[005: origami\_cherry\_blossoms] Origami cherry blossoms that fall when you pick the first petal :cherry-blossom-cowboy: Make pedals collide with physical environment using depth mesh.
  \item[006: wall\_aquarium] Create a new app, When I click a detected mesh of the wall, a virtual fish tank should be created and carved into that detective mesh. The final result should looks like there is a fish tank which inside the wall.
  \item[007: cat\_platformer] Create a cat-themed super mario alike game.
  \item[008: planetary\_gearbox] Create a model of a planetary gearbox and display it in the app. Animate the model to demonstrate how the device works. Add teeth to the gears in the model to make the rotation of the gears easier to see.
  \item[009: guitar\_tab\_tutor] Create an app with a model of a guitar that shows dots on strings and frets that correlate to tablature and show you how to play a song. Use Jimi Hendrix little wing as an example.
  \item[010: hand\_fire] Create a WebXR application called "Hand Fire" using three.js and xrblocks that tracks both hands to emit distinct particle-based fire effects, blue fire for the left hand and orange fire for the right. Implement a custom CPU-driven particle system with a shader for the visual effect, and include gesture logic where pinching makes the fire follow the fingertips, while a "thumbs up" gesture freezes the fire in mid-air at its current location. Additionally, provide a spatial UI panel with an "Ignite Hands" toggle button that forces the fire to appear immediately (floating in front of the camera if hands are not detected, or snapping to the palm center if they are), allowing for both manual and gesture-based control of the effects.
  \item[011: neon\_dodge\_arena] Create a high-intensity neon dodge arena. The visuals should be dominated by glowing primitives, particle trails, and chromatic aberration. The Hero: A soft-glow orb that leaves a light trail. The Threat: Procedural bullet patterns that pulse to a rhythmic tempo. The Flow: Add a "Bullet Time" meter for tactical slowing and a combo counter that ticks up when players "near-miss" projectiles. The Experience: The game should launch into a self-playing cinematic demo (Attract Mode) that seamlessly transitions to gameplay upon player input. Decoration a variety of spring festival lanterns in my physical environment, when I pinch, spawn a new lantern
  \item[012: origami\_bird] Make an origami bird that flies around the room for a few seconds and then lands on my hand. When I move my hand, it flies away and repeats.
  \item[013: voxel\_parthenon] generate a voxel Parthenon building
  \item[014: bird\_quiz\_game] create a bird quiz game in 3D
  \item[015: furniture\_placement] Build a simple 3D scene using basic geometry and orbit controls that features a furniture placement system and a dynamic day-night lighting cycle.
  \item[016: alpine\_shiba\_cabin] A photorealistic alpine meadow with wildflowers. Among the evergreen pine trees is a rustic log cabin with a front porch, with a shiba inu outside the cabin in front of the user.
  \item[017: stickman\_sketch] Develop an interactive AR/3D application where users can sketch a stickman in three-dimensional space, featuring a "Finish" toggle that converts the static drawing into a rigged, physics-based character capable of procedural animation and reactive haptics (such as recoiling or stepping back) upon collision with the user's hand.
  \item[018: underwater\_fish\_scene] Create a photorealistic underwater scene featuring schools of fish swimming naturally through a complex environment of coral and rocks, utilizing advanced depth rendering for realistic environmental occlusion.
  \item[019: procedural\_vine\_growth] Generate a procedural vine growth simulation featuring recursive branching, dynamic tendril curling with physics-based wall-climbing, and time-remapped triggers for leaf sprouting and blooming flowers along an organic spreading path.
  \item[020: voxel\_garden] Create an app with XR Blocks that is delightful and fun. Users can aim with either hand to raycast their cursor against the environment, using depth perception. When they do a pinch gesture, a plant starts growing from the place they aimed at. They can do it many times to create many plants all around their environment. Every time they pinch, a different plant should appear. It can be a palm tree, and oak tree, a cactus, a baobab, a rose. All the plants should be made out of voxels. Their growth should be animated, and it should take 5 seconds for the plant to reach its full size. Plants at their full size should be no more than 2 meters tall so that they fit in indoor environments.
  \item[021: waterfall\_simulation] Develop a waterfall simulation featuring cascading particles, mist spray against rugged rock formations, realistic pool splashing, and a refractive rainbow effect within the mist.
  \item[022: weather\_prototype] Make an immersive weather prototype that can switch between sunny, rainy, thunderstorm, and fog with two buttons
  \item[023: edm\_concert] edm concert environment with blurry lights around
  \item[024: flower\_watering] Create an interactive scene where user can pour the water and water the flower when the hand pinch gesture is detected.
  \item[025: four\_bar\_linkage] Create a scene to show a basic four-bar linkage
  \item[026: spatial\_invaders] Create a space invaders game. Spatial Playfield: A grid of animated "alien" meshes that descend toward the user. Controller Interaction: Shooting projectiles from your hands or mouse click
  \item[027: spatial\_pacman] Spatial Pacman. There should be a spatial grid based on depth of the space the user is in, and the movement should happen with user gestures.
  \item[028: voxel\_tower\_defense] Create an XR tower defense game where I can place voxel towers on a real table and the towers will attack randomly created voxel monsters.
  \item[029: fiery\_origami\_horse] build a fiery horse out of origami that dances to audio from my microphone
  \item[030: chill\_cats] Create a relaxing experience where cats are wandering around in my space around me and just hanging out chill
  \item[031: toilet\_paper\_rain] Tons of toilet papers falling from your room ceiling, whenever you look up. Toilet paper rolls should land on the floor. The floor and furnitures should have colliders
  \item[032: chunky\_holographic\_horse] Create a 3D scene loading a standard horse GLB model. Set the horse to a continuous gallop loop. Make the horse look slightly fat/chunky using vertex manipulation and apply a glowing holographic shader skin. Add motion trail particles behind the horse to emphasize speed.
  \item[033: snowboard\_simulator] make a mini game that can help me practice snowboarding, simulate an indoor snow resort
  \item[034: schrodingers\_cat] An aesthetically pleasing depiction of \\ Schrödinger's cat in XR. Finger pinch makes a cat (detailed 3D model) go into the box. Approaching the box within 50cm makes the box become two that move to the left and right and the boxes front wall becomes transparent. You see both versions of the cat inside (dead and alive), demonstrating the Quantum mechanics. When you pinch again, one of the states becomes reality. The box opens and you see it either alive or dead. With another pinch you can start again.
  \item[035: yellowstone\_terrain] render and visualize the terrain of the yellowstone national park and add educational markers to famous spot, when clicking on them, show panels to introduce the knowledge for that terrain / feature
  \item[036: xr\_pomodoro] make a productivity app of a tomato clock in xr that I can place on my desk
  \item[037: monster\_escape\_room] create an app where user can play escape room, and user can choose to generate the monster he wants. The monster can be a horse gradually evolving into a horse man.
  \item[038: snow\_white] Create a storytelling scene of snow white.
  \item[039: butterfly\_catching] Create an XR butterfly catching game where I hold a net and can use it to catch butterflies
  \item[040: tool\_shopping] Build an XR shopping game where I can pick up from tools like a drill, hammer, and knife. Each item should have a price tag and have physics so when I let go they drop to the floor.
  \item[041: marble\_run] Create a Mixed-Reality Marble Run. The user places tracks to lead a marble from a start point to a goal. Use the depth reticle to anchor "start" and "end" points on different parts.  Make the marble bounce off your actual palms to keep it from falling.
  \item[042: cat\_fireplace\_fireflies] Create a fireplace with a furry cat playing with fireflies nearby
  \item[043: ar\_math\_graph] Open camera, when I give Gemini a math problem, I can watch it visualizes everything in a 3D AR graph
  \item[044: spatial\_fish] Fish swimming across the space
  \item[045: voxel\_garden] Create an app with XR Blocks that is delightful and fun.

Users can aim with either hand to raycast their cursor against the environment, using depth perception. When they do a pinch gesture, a plant starts growing from the place they aimed at. They can do it many times to create many plants all around their environment.

Every time they pinch, a group of plants should appear. In each group, there should always be a big plant in the middle, surrounded by between 3 and 5 smaller flowers in a 30cm radius around the big plant. The big plant can be either a sunflower, a fiddle-leaf fig, or a cactus. These big plants should be 1 meter tall. The other flowers should be smaller, no taller than 20cm when at their full size. They can be flowers of different colors, like roses (pink or red), tulips (reds, yellows, whites, pinks, oranges or deep purples), or dandelions.
There should also be a little bit of grass around the flowers. All the plants should be made out of voxels. Their growth should be animated, and it should take 5 seconds for the plant to reach its full size. It is important that the animations work properly to create delight. Each plant within a group should grow individually. All plants should always be oriented upwards, and not based on the normal of the depth map.

If they maintain their pinch and drag, new small plants should be planted along the path they they are drawing with their cursor.

Every time a group of plants is added, there should be a few beautiful piano notes being played  (one per new plant, creating a beautiful piano chord) as the plants grow to illustrate the beauty of growing nature. If pinching and dragging to plant flowers along a path, there should only be a note every 3 flowers to not make the sound too overwhelming.

Once plants have been added, there should be some voxel insects (bees and ladybugs) flying around the area. There should be as many insects as there are groups of plants. One new insect should be spawn in the area every time there is a new group of plants. They should start being attracted by the plants and land on them. After a bit, they should take off and fly to a different plant. They should stay keep doing this and visit random plants this way forever.

Do not show a laser pointer coming out of my hands.
Make sure you import BufferGeometryUtils as a namespace to avoid import errors.
  \item[046: plant\_doctor] Plant doctor: AI expert can help with care, routine, mock watering tracking with plants you point at.
  \item[047: xr\_home\_decorator] XR home decorator: add virtual picture frames, 3D assets, mini-games around your home. Use depth + AI to suggest locations and widgets.
  \item[048: social\_home\_cinema] use your living room geometry and digitize/retexture it and decorate it. Invite people to hang out and watch movie together.
  \item[049: breathing\_exercise] breathing exercise with 3D visuals
  \item[050: matrix\_mesh] Matrix Effect with meshes
  \item[051: art\_exhibition] Curate and arrange an art exhibition in the XR world
  \item[052: object\_capture] Scan and copy real world objects into xr world
  \item[053: gemini\_tutorial] Gemini becomes a flying light ball, teach users how to interact with the app.
  \item[054: drawing\_guess] let gemini guess what I’m drawing
  \item[055: ping\_pong] Two player ping pong
  \item[056: sticky\_notes] Sticky Notes around the house
  \item[057: yoga\_instructor] Real time yoga instructor in your room
  \item[058: ar\_weather] “What will the weather be like tomorrow?” - show visual effects in AR that show rain, clouds, etc.
  \item[059: sonar\_vision] “Daredevil” sonar vision simulator
  \item[060: superpower\_hands] Create a delightful app that gives me super powers.

The room should be filled with pixie dust, and pinching with either of my hands should attract particles toward it, and make each particle individually disappears once it reaches it. Particles that have not reached the hand yet should remain visible. The goal is to clear the room from all the pixie dust.

The dust should be made of very tiny shiny yellow dots, made from a billboarded quad with a radial gradient to make it look like a little sparkle. The particles are floating in mid air with a slow floating motion, and there should be thousands of them in the room. There should be some physics applied to the particles, so if I start pinching, they are attracted by my hand, and if I release the pinch, the still carry some momentum and slow down gradually.

Pinching should attract the particles that are within a cone that has its apex at the position of my hand, and its main axis going toward the opposite direction of the camera (my head). This allows me to attract particles that are far away by aiming toward them with me hand.

For each particle that reaches the hand and disappears, there should be a large flash a the position of my pinch. This will make clearing the particles more gratifying. The flash is made of a quad, with a radial gradient, the same color as the particle that just got cleared.

The particles should be occluded by the environment and by my hand using the depth sensing capabilities.

There should be a floating text 1.5m away from the user at start time that says "Aim and pinch with your hand to attract and clear the dust."
Below, there should be a big percentage that shows the percentage of particles that have already been cleared.

Once all the particles have been cleared, there should be fireworks exploding all around the room. They should keep exploding in random locations forever so that the celebration never stops, until the user closes the app.
\end{description}

\section{Application Scenarios Prompts}
\label{sec:application_scnarios_prompts}

\begin{description}
  \item[AS1: euler\_visualization] Visualize Euler's theorem in geometry. Explain vertices, edges, and facets concepts with highlighting using different examples.
  \item[AS2: physics\_lab] create an interactive physics experiment: given different objects on each side of the scale, use different weights (with labels on them) to balance the scale.
  \item[AS3: immersive\_chemistry] create an interactive chemistry lab that users can use hands pinch to ignite and observe three experiments: Ignite methane in air and place a dry, cold beaker over the flame: the flame is pale blue, and liquid droplets form on the inner wall of the beaker. Ignite ethylene in air: the flame is bright, black smoke is produced, and heat is released. Ignite acetylene in air: the flame is bright, thick smoke is produced, and heat is released.
  \item[AS4: xr\_sports] Let me play volleyball with hands and collide with my environment. Volleyballs are textured and launched from a red ring slowly and easier to bounce with the hand.
  \item[AS5: chrome\_dino] create the Chrome Dino game in xr. Dino is voxelized in front of the user, letting every cactus rushing towards the user on a semi transparent lane. Add audio.
\end{description}

\section{System Prompts of \thesystem}
\label{sec:systemprompts}

\begin{lstlisting}
# Role
Act as an expert Creative Technologist specialized in three.js, WebXR, and the **XR Blocks (xrblocks)** library.

# Context
You are authoring single-file WebXR experiences. You must strictly adhere to the XR Blocks framework architecture.

# User Request
Create the app following user request: <my_xr_experience>, following the XR Blocks examples.

# Engineering Guidelines (Strict)

1. **Architecture (Single File):**
   - Output a SINGLE `index.html` file.
   - Logic must be inside a class extending `xb.Script` within `<script type="module">`.
   - XR Blocks handles XR sessions, window resizing, and rendering loop.

2. **Dependency Management (Critical):**
   - Use the specific versions below. Do NOT hallucinate newer versions.
   - Only include imports required for the specific request (e.g., do not import TensorFlow unless using gestures).
   - **Reference Map:**
     // omitted

3. **Coding Standards:**
   - **Class Structure:** logic must be inside `class MyScript extends xb.Script`.
   - **Lifecycle:** Use `init()` for setup and `update()` for the loop. Do NOT use `requestAnimationFrame` manually; `xb.Script` handles this.
   - **Interactions:** Use `onSelectStart`, `onSelectEnd`, `onSelecting` methods within the class.
   - **Coordinates:** `y` is up. `z` is forward/backward. Initialize objects at `z = -this.user.objectDistance`.
   - **Spatial UI:** Use XR Blocks 3D UI instead of 2D overlay divs.

4. **Specific XR Features:**
   - **Text:** Use `troika-three-text` (Pattern: `1_ui`).
   - **Models:** Wrap 3D models in `xb.ModelViewer`.
   - **AI:** If using Gemini/AI, use const API_KEY = "", and link the API key to the `ai` module so Gemini Canvas auto-populates it.

5. **Planning:**
   - Before generating code, briefly outline the `xb.Script` class structure, member variables needed, and the `init()` vs `update()` logic flow.

# Reference Examples
Here are some XRBlocks **templates, samples, demos, and gallery examples**...

\end{lstlisting}

This shows the additional parts other than \xrblocks examples.
The latest full prompts with XR Blocks examples are shared in \prompturl, 
where the prompt for this work is shared in \href{https://github.com/xrblocks/xrblocks.github.io/blob/c06978cb7df8aac84a126ef97c0cc0c15aceb0d9/build/prompts.txt}{this Github commit}.

\section{Example snippets of XR Blocks}

\subsection{Basic Pinch}
\label{sec:pinch}

\begin{jscode}
import * as THREE from 'three';
import * as xb from 'xrblocks';

/**
 * A basic example of XRBlocks to render a cylinder and pinch to change color.
 */
class MainScript extends xb.Script {
  init() {
    this.add(new THREE.HemisphereLight(0xffffff, 0x666666, /*intensity=*/ 3));

    const geometry = new THREE.CylinderGeometry(
      0.2,
      0.2,
      0.4,
      /*segments=*/ 32
    );
    const material = new THREE.MeshPhongMaterial({
      color: 0xffffff,
      transparent: true,
      opacity: 0.8,
    });
    this.player = new THREE.Mesh(geometry, material);
    this.player.position.set(0, xb.user.height - 0.5, -xb.user.objectDistance);
    this.add(this.player);
  }

  /**
   * Changes the color of the mesh on a pinch in XR.
   * @param {SelectEvent} event event.target holds controler / hand data.
   */
  onSelectEnd(event) {
    this.player.material.color.set(Math.random() * 0xffffff);
  }

  /**
   * Changes the color of the mesh to blue during pinching in XR.
   * @param {SelectEvent} event The controller / hand event.
   */
  onSelecting(event) {
    this.player.material.color.set(0x66ccff);
  }

  dispose() {
    this.player.geometry.dispose();
    this.player.material.dispose();
  }
}

document.addEventListener('DOMContentLoaded', function () {
  xb.add(new MainScript());
  xb.init(new xb.Options());
});
\end{jscode}

\subsection{Basic Camera + VLM}

\begin{jscode}
export class PoemGenerator extends xb.Script {
  constructor() {
    super();
    this.panel = null;
    this.isProcessing = false;
    this.responseDisplay = null;
    this.button = null;
    this.loadingSpinner = null;
    this.poemGenerated = false;
    this.hasError = false;
  }

  init() {
    this.ai = xb.core.ai;
    this.deviceCamera = xb.core.deviceCamera;

    this.createPanel();
  }

  createPanel() {
    this.panel = new xb.SpatialPanel({
      width: 2.0,
      height: 1.25,
      backgroundColor: '#1a1a1abb',
    });
    this.panel.position.set(0, 1.6, -2);
    this.add(this.panel);

    const grid = this.panel.addGrid();

    const responseRow = grid.addRow({weight: 0.9});
    this.responseDisplay = new xb.ScrollingTroikaTextView({
      text: '',
      fontSize: 0.03,
    });
    responseRow.add(this.responseDisplay);

    const buttonRow = grid.addRow({weight: 0.2});

    const videoRow = grid.addRow({weight: 0.7});
    this.videoView = new xb.VideoView({
      width: 1.0,
      height: 1.0,
      mode: 'stretch',
    });
    videoRow.add(this.videoView);

    const buttonPanel = buttonRow.addPanel({
      backgroundColor: '#00000000',
      showEdge: false,
    });
    this.button = buttonPanel.addGrid().addIconButton({
      text: 'photo_camera',
      fontSize: 0.6,
      backgroundColor: '#FFFFFF',
      defaultOpacity: 0.2,
      hoverOpacity: 0.8,
      selectedOpacity: 1.0,
    });
    this.button.onTriggered = () => this.captureAndGeneratePoem();

    if (this.deviceCamera) {
      this.videoView.load(this.deviceCamera);
    }

    // Create 3D loading spinner
    this.loadingSpinner = new TorusKnotLoadingAnimation();
    this.panel.add(this.loadingSpinner);
  }

  async captureAndGeneratePoem() {
    if (this.isProcessing || !this.ai?.isAvailable()) return;
    this.isProcessing = true;
    this.hasError = false;
    if (this.button) this.button.disabled = true;
    if (this.loadingSpinner) this.loadingSpinner.show();

    try {
      const snapshot = await this.deviceCamera.getSnapshot({
        outputFormat: 'base64',
      });
      if (!snapshot) {
        throw new Error('Failed to capture video snapshot.');
      }
      const {strippedBase64, mimeType} = xb.parseBase64DataURL(snapshot);
      const image = {inlineData: {mimeType: mimeType, data: strippedBase64}};
      const question =
        'Can you write a 12 lined, lighthearted poem about what you see?';
      const parts = [image, {text: question}];

      const response = await this.ai.query({
        type: 'multiPart',
        parts: parts,
      });
      this.responseDisplay.setText(`${response.text}\n\n`);
      this.poemGenerated = true;
    } catch (error) {
      this.responseDisplay.setText(`Error: ${error.message}\n\n`);
      this.hasError = true;
    } finally {
      this.isProcessing = false;
      // Re-enable button
      if (this.button) {
        this.button.disabled = false;
        this.button.update();
      }
    }
  }

  update() {
    if (this.loadingSpinner) {
      this.loadingSpinner.update(this.isProcessing);
    }
  }
}
\end{jscode}

\subsection{XR Object Recognition}

\begin{jscode}
/**
 * A basic example of using the XR Blocks SDK's world component to detect
 * objects in the real world using Gemini.
 */
class MainScript extends xb.Script {
  init() {
    this.add(new THREE.HemisphereLight(0xffffff, 0x666666, /*intensity=*/ 3));
  }

  /**
   * Runs object detection on select (click in simulator, pinch in XR device).
   * The results of the detection are automatically handled by the
   * ObjectDetector, which will create debug visuals for each detected object.
   * This behavior is enabled when `showDebugVisualizations` is set to true.
   * @param {XRInputSourceEvent} event event.target holds controller or hand
   * data.
   */
  async onSelectEnd() {
    console.log('Running object detection...');
    const detectedObjects = await xb.world.objects.runDetection();

    // `detectedObjects` is an array of THREE.Object3D instances, each
    // representing a detected object. These objects contain the 3D world
    // position and other metadata returned by the detection model.
    if (detectedObjects.length > 0) {
      console.log('Detected objects:', detectedObjects);
    } else {
      console.log('No objects detected.');
    }
  }
}

document.addEventListener('DOMContentLoaded', function () {
  const options = new xb.Options();

  // AI is required for the object detection backend. Please set your
  options.enableAI();

  // Enable the environment camera to provide the video feed to the AI module.
  options.enableCamera('environment');

  // Depth is required to project the 2D detections from the AI module into 3D.
  options.enableDepth();

  // Enable the object detection feature and its debug visualizations.
  options.world.enableObjectDetection();
  options.world.objects.showDebugVisualizations = true;

  xb.add(new MainScript());
  xb.init(options);
});
\end{jscode}

\end{document}